\documentclass[]{aa} %for a referee version

\usepackage[utf8]{inputenc}
\usepackage{graphicx}
\usepackage{txfonts}
\usepackage{natbib}
\usepackage{float}

\usepackage{setspace} %\singlespacing

\usepackage{amsmath}
\usepackage{graphicx}

\bibpunct{(}{)}{;}{a}{}{,}
\graphicspath{{figures/}}
\title{Generation of solar chromosphere heating and coronal outflows by two-fluid waves}
\authorrunning {M.~Pelekhata et al.}
\titlerunning {Two-fluid Alfv\'en-magnetoacoustic waves in the solar atmosphere}
   \author{M.~Pelekhata\inst{1}
          \and
          K.~Murawski\inst{1}
          \and 
          S.~Poedts\inst{2,1}
          }

   \institute{Institute of Physics, University of M. Curie-Sk{\l}odowska, 
              Pl.\ M.\ Curie-Sk{\l}odowskiej 1, 20-031 Lublin, Poland\\
         \and
            Centre for Mathematical Plasma Astrophysics / Department of Mathematics, KU Leuven, Celestijnenlaan 200B, 3001 Leuven, Belgium\\
             }

\begin{document}

\abstract
%{Context:} 
{
It is known that Alfv\'en and magnetoacoustic waves both contribute to the heating of the solar chromosphere and drive plasma outflows. In both cases, the thermalization of the wave energy occurs due to ion-neutral collisions, but the obtained rates of plasma heating cannot explain the observational data. The same is true for the magnitudes of the outflows. 
}
%{Aims:} 
{
The aim of the present paper is to reexamine two-fluid modeling of Alfv\'en and magnetoacoustic waves in the partially ionized solar chromosphere. We attempt to detect variations in the ion temperature and vertical plasma flows for different wave combinations. 
}
%{Methods:} 
{
We performed numerical simulations of the generation and evolution of coupled Alfv\'en and magnetoacoustic waves using the JOANNA code, which solves the two-fluid equations for ions (protons)+electrons and neutrals (hydrogen atoms), coupled by collision terms.
}
%{Results:} 
{
We confirm that the damping of impulsively generated small-amplitude waves negligibly affects the chromosphere temperature and generates only slow plasma flows. In contrast, waves generated by large-amplitude pulses significantly increase the chromospheric temperature and result in faster plasma outflows. The maximum heating occurs when the pulse is launched from the center of the photosphere, and the magnitude of the related plasma flows increases with the amplitude of the pulse.
}
%{Conclusions:} 
{
Large-amplitude coupled two-fluid Alfv\'en and magnetoacoustic waves can significantly contribute to the heating of the solar chromosphere and to the generation of plasma outflows. 
}

\maketitle

\section{Introduction}

The solar atmosphere is a gravitationally stratified and magnetically structured medium, in which the temperature, mass density, gas pressure, and ionization degree vary with height. As a result of this, the atmosphere can be theoretically divided into the following layers with different physical characteristics: the photosphere, the chromosphere, the transition region, and the solar corona. The bottom of the photosphere is located at the top of the convection zone and it extends up to $500\;$km in height. The next layer, called the chromosphere, develops up to the level of about $2500\;$km. The corona caps the chromosphere and spreads out into the solar wind over a distance of about 2-3 solar radii for the low corona, and even up to 20 solar radii in some other models. Between the chromosphere and the corona, a narrow plasma layer of only $100-200\;$km thick, called the transition region, settles in. The most significant feature for the present paper is the temperature variation between (and in) these layers \citep[e.g.,][]{2008ApJS..175..229A}
as it leads to a strongly varying degree of ionization in the solar atmosphere \citep[e.g.,][]{2017PPCF...59a4038K}.  

At the bottom of the photosphere, the temperature is only about $5600\;$K. Then, it gradually falls off with the height to its minimum, of about $4300\;$K. This temperature minimum level is located at about $100\;$km above the photosphere \citep{1976ASSL...53.....A}. Higher up, the temperature rises again, first gradually in the low chromosphere, and then the temperature increase accelerates from the high chromosphere up to the transition region where, on average, the temperature is $10^{4} - 10^{5}\;$K. In the transition region, the temperature abruptly increases and it reaches values of $1-3$ million K in the solar corona \citep{2005SoPh..228..339A}. 
The reason for this temperature rise with height remains one of the major problems of heliophysics \citep{1974SoPh...35..451U, 2010LRSP....7....4O}. 

The ionization degree is defined as fraction of particles that are ionized. It directly depends on the plasma temperature; in other words, the lower the temperature, the lower the ionization degree. A low ionization degree means that most of the matter is not ionized, with many atoms being able to hold their electrons. As a result of its enormous temperature, the solar corona is fully ionized \citep{2005psci.book.....A}. In contrast, the lower layers of the solar atmosphere are only partially ionized \citep{2003ASPC..286..419A}. In the upper photosphere, at the temperature minimum, the ionization degree is only about $10^{-4}$, which means that there is only one ion per about $10^4$ neutrals. In the chromosphere the ionization degree grows with height, which motivates and justifies the use of the two-fluid model of the solar atmosphere. 

In the two-fluid model used in the present paper, ions+electrons and neutrals are treated as two separate fluids. Here, only a neutral hydrogen atom was considered, but a substantial amount of neutral helium atoms may also be present in the plasma under the condition of a particular temperature (about $10^{4}-4\cdot10^{4}\;$K) and ionization. In \citet{2011A&A...534A..93Z}, the importance of ions with neutral helium atom collisions in chromospheric spicules and in prominence-corona transition regions was shown. The presence of neutral helium would significantly affect the damping of Alfv\'en waves in comparison to the damping due to only neutral hydrogen.

Different ideas have been developed to explain the sudden temperature increase at the transition region. Some of them involve Alfv\'en waves, claiming that they can be a part of the solution to this problem \citep{1956MNRAS.116..314P, 1961ApJ...134..347O}. For instance, \citet{2016ApJ...819L..24Y} revealed that Alfv\'en waves may carry a sufficient amount of energy to heat the corona. \citet{2004A&A...427.1055E} proposed that ion-neutral collisions cause damping of Alfv\'en waves, which in turn exerts an impact on the increase in the chromospheric temperature \citep{2005A&A...442.1091L, 2011ApJ...735...45G, 2013ApJ...777...53T, 2013A&A...549A.113Z, 2016ApJ...817...94A, 2016AGUFMSH21E2565S, 2017ApJ...840...20S}. \citet{2018SSRv..214...58B} showed that ambipolar diffusion leads to chromospheric plasma heating, and \citet{2013A&A...549A.113Z} proved that the collisional damping of Alfv\'en waves is actually significant in the chromosphere. The mechanisms of wave damping due to the ion-neutral collisions were investigated by \citet{2001ApJ...558..859D}. \citet{2013A&A...549A.113Z} and \citet{2017ApJ...840...20S} proposed that Alfv\'en waves that are formed in the photosphere, with wave periods of a few seconds might not reach the solar corona because they are efficiently damped by ion-neutral collisions in the upper chromosphere. The wave damping depends both on the strength of the magnetic field and on the wave period; the stronger the field, the lower the damping, and larger (in comparison to the collision time) period waves are more weakly damped (see \citet{2011JGRA..116.9104S}).

Actually, \cite{1946NW.....33..118B} and \cite{1948ApJ...107....1S} first suggested that acoustic waves may be responsible for chromosphere heating. Afterward, this topic was studied many times and these investigations revealed that these waves are indeed able to heat the chromosphere \citep{1995ApJ...440L..29C, 2003ASPC..286..363U, 2017ApJ...849...62N, 2019ApJ...878...81K}. \citet{2021A&A...652A..88K} showed that the properties of magnetoacoustic waves depend on the configuration of the ambient magnetic field. Also, the problem of the damping of these waves was investigated by \citet{2021arXiv211204995P} and \citet{2021A&A...646A.155D}. \citet{2019ApJ...878...81K} showed that acoustic waves thermalize their energy by ion-neutral collisions in the chromosphere. \citet{2019A&A...627A..25P} extended the model of \citet{2019ApJ...878...81K} on magnetoacoustic waves.

Numerous papers reported on the presence of Alfv\'en waves \citep{1942Natur.150..405A, 2007Sci...317.1192T, 2017NatSR...743147S, 2021ApJ...907...16B} and magnetoacoustic waves \citep{1946NW.....33..118B, 1948ApJ...107....1S} in the solar atmosphere. Alfv\'en waves are transverse magnetohydrodynamic (MHD) waves that can only travel along magnetic field lines. When they pass by, they alter the azimuthal components of the magnetic field and the plasma velocity, that is to say the components within the flux surfaces but perpendicular to the magnetic field. In the linear limit, Alfv\'en waves do not modify the gas pressure nor the mass density \citep{2005LRSP....2....3N}, so they are incompressible. Some of them arise from the dense photosphere and occasionally are reflected into the photosphere; however, some can reach the chromosphere or even the solar corona \citep{2010A&A...518A..37M}. Many observational data confirm the presence of Alfv\'en waves in the chromosphere and corona \citep{2008ApJ...687L.131B, 2009Sci...323.1582J, 2009A&A...507L...9W}. Nevertheless, \citet{2008ApJ...676L..73V} proposed that some of these waves can be interpreted as fast magnetoacoustic waves. Also, it was shown that Alfv\'en waves can turn out to be nonlinear in the chromosphere. These kind of waves can drive magnetoacoustic waves by ponderomotive force  \citep{2009ApJ...700L..39V, 2010ApJ...710.1857M}. Magnetoacoustic waves are associated with perturbations of gas pressure and mass density. These waves can be divided into slow and fast waves. Fast magnetoacoustic waves are driven by perturbations in the gas and magnetic pressures, which act in phase. For slow magnetoacoustic waves, the perturbations in gas and magnetic pressures work in antiphase. In a strongly magnetized medium, slow waves cannot travel perpendicular to the direction of the magnetic field, and fast waves move quasi-isotropically \citep{2021arXiv211214486S}. 

Recently, a potential contribution of two-fluid Alfv\'en waves to the heating of the solar chromosphere and the generation of plasma outflows was investigated by \citet{2021A&A...652A.114P}. It was found that a significant temperature increase was only observed for large amplitudes of the initial pulse and that these waves can drive plasma outflows that, higher up, may originate the solar wind. It was specified that the maximum heating occurs for a pulse launched from the middle of the photosphere, mainly from $y\approx0.3\;$Mm, and with the maximum pulse amplitude $A=10\;\text{km}\cdot\text{s}^{-1}$. 
In the parallel research performed by \citet{2021A&A...652A.124N}, the effect of magnetoacoustic waves was studied in a similar framework, to show that these waves can also increase the chromospheric temperature and induce plasma outflows. In particular, \citet{2021A&A...652A.124N} found that the heating rate grows with the initial pulse amplitude and with its width. In contrast, raising the altitude at which the pulse is launched from results in opposite effects, mainly in a local temperature reduction and slower plasma outflows. 

In the case of Alfv\'en \citep{2021A&A...652A.114P} and magnetoacoustic \citep{2021A&A...652A.124N} waves, heating of the chromosphere took place due to ion-neutral collisions. Both studies were performed using two-fluid and magnetohydrostatic equilibrium models. Considering this, the present paper aims to study a combination of Alfv\'en and magnetoacoustic waves in the solar atmosphere. More precisely, this paper examines the propagation of impulsively generated Alfv\'en and magnetoacoustic waves in the context of plasma heating and the generation of plasma flows.

The organization of the remainder of this paper is as follows. In Sect.~2 the two-fluid equations are presented, as well as the background equilibrium model of the solar atmosphere, and the impulsive perturbations that were applied in the numerical simulations. In Sect.~3, the results of the numerical simulations are presented, and Sect.~4 contains a discussion and summary of the results of the numerical experiments performed, and the conclusions that can be drawn from them.
\section{Physical model}

A gravitationally stratified and partially ionized solar atmosphere is used to model the Sun's lower atmospheric layers. Due to the substantial presence of neutral particles in these lower layers \citep{2014PhPl...21i2901K}, a two-fluid plasma model is used. For the sake of simplicity, ions and electrons are represented by a single ion-electron fluid, whereas neutrals are described as a second fluid.  
These two fluids each have their own mass density, flow velocity, and gas pressure, and interaction between them is ensued via ion-neutral collisions.

\subsection{Two-fluid equations}
The evolution of the chosen Sun's atmospheric area in this model is described by the two-fluid equations \citep{2011A&A...529A..82Z, 2012ApJ...760..109L, 2013A&A...549A.113Z, 2018SSRv..214...58B, 2018ApJ...856...16M}. The two-fluid equations are a combination of the MHD equations for charges and the Navier-Stokes equations for the neutrals. These equations can be written in the following way \citep{2018SSRv..214...58B, 2015hsa8.conf..677K}:
\begin{eqnarray}
    &&\frac{\partial \varrho_{\rm ie}} {\partial t}+\nabla\cdot(\varrho_{\rm ie} \mathbf{V}_{\rm ie}) = 0\,, 
    \label{eq:ion_continuity} \\
    &&\frac{\partial \varrho_{\rm n}}{\partial t}+ \nabla\cdot(\varrho_{\rm n} \mathbf{V}_{\rm n}) = 0\,, 
    \label{eq:neutral_continuity} \\
    &&\frac{\partial (\varrho_{\rm ie} \mathbf{V}_{\rm ie})}{\partial t} + \nabla \cdot (\varrho_{\rm ie} \mathbf{V}_{\rm ie} \mathbf{V}_{\rm ie} + p_{\rm i e}\mathbf{I}) = \\ \nonumber && \varrho_{\rm ie} \mathbf{g} + \frac{1}{\mu}(\nabla \times \mathbf{B}) \times \mathbf{B} - v_{\rm in}\varrho_{\rm ie}({\bf V_{\rm ie}}-{\bf V_{\rm n}})\,,
    \label{eq:ion_momentum}\\
    &&\frac{\partial (\varrho_{\rm n} \mathbf{V}_{\rm n})}{\partial t} + \nabla \cdot (\varrho_{\rm n} \mathbf{V}_{\rm n} \mathbf{V}_{\rm n} + p_{\rm n} \mathbf{I})  =  \\ \nonumber && \varrho_{\rm n} \mathbf{g} + v_{\rm in}\varrho_{\rm ie}({\bf V_{\rm ie}}-{\bf V_{\rm n}})\,,
    \label{eq:neutral_momentum} \\
    &&\frac{\partial E_{\rm ie}}{\partial t} + \nabla\cdot\left[\left(E_{\rm ie}+p_{\rm i e} + \frac{\mathbf{B}^2}{2\mu} \right)\mathbf{V}_{\rm ie}-\frac{\mathbf{B}}{\mu}(\mathbf{V}_{\rm ie}\cdot \mathbf{B})\right] = \\ \nonumber && (\varrho_{\rm ie} \mathbf{g} + v_{\rm in}\varrho_{\rm ie}({\bf V_{\rm ie}}-{\bf V_{\rm n}})) \cdot \mathbf{V}_{\rm ie} + Q_{\rm ie}\,, 
    \label{eq:ion_energy} \\
   &&\frac{\partial E_{\rm n}}{\partial t}+\nabla\cdot[(E_{\rm n}+p_{\rm n})\mathbf{V}_{\rm n}] = \\ \nonumber && (\varrho_{\rm n} \mathbf{g} + v_{\rm in}\varrho_{\rm ie}({\bf V_{\rm ie}}-{\bf V_{\rm n}}))\cdot\mathbf{V}_{\rm n} + Q_{\rm n}\,,
    \label{eq:neutral_energy} \\
\hbox{with}&& \nonumber\\
    &&\frac{\partial \mathbf{B}}{\partial t} = \nabla \times (\mathbf{V_{\rm ie} \times }\mathbf{B}),
    \hspace{6mm}
    \nabla \cdot{\mathbf B} = 0\,, \\
    &&E_{\rm ie} = \frac{\varrho_{\rm ie}\mathbf{V}_{\rm ie}^2}{2} + \frac{p_{\rm i e}}{\gamma -1 } + \frac{{\mathbf B}^2}{2\mu}\,, 
    \hspace{0.5mm} 
%\\
%
    E_{\rm n} = \frac{\varrho_{\rm n}\mathbf{V}_{\rm n}^2}{2} + \frac{p_{\rm n}}{\gamma -1 }\,,  \\
\hbox{where:}&& \nonumber\\
    &&p_{\rm ie}=\frac{k_{\rm B}}{m_{\rm ie}}\varrho_{\rm ie} T_{\rm ie}\,,
    \quad
   p_{\rm n}=\frac{k_{\rm B}}{m_{\rm n}}\varrho_{\rm n} T_{\rm n}\,, \\
    &&Q_{\rm ie}=\frac{1}{2}\nu_{\rm in}\varrho_{\rm ie}({\bf V_{\rm ie}}-{\bf V_{\rm n}})^2 
    -
    %\frac{1}{\gamma - 1}
    3
    \frac{\nu_{\rm in}\varrho_{\rm ie}k_{\rm B}}
    {m_{\rm ie}+m_{\rm n}}(T_{\rm ie}-T_{\rm n})\,,  \\
    &&Q_{\rm n}=\frac{1}{2}\nu_{\rm in}\varrho_{\rm ie}({\bf V_{\rm ie}}-{\bf V_{\rm n}})^2
    -
    %\frac{1}{\gamma - 1}
    3
    \frac{\nu_{\rm in}\varrho_{\rm ie}k_{\rm B}}{m_{\rm ie}+m_{\rm n}}(T_{\rm n}-T_{\rm ie})\,.
\end{eqnarray}
Here, the indices $\rm_{i, e, n}$ correspond to ions (protons),  electrons, and neutrals (hydrogen atoms), respectively. Therefore, ${\bf V}_{\rm i}$ and ${\bf V}_{\rm n}$ are, respectively, ion and neutral velocities, ${\bf B}$ is the magnetic field, ${\bf I}$ indicates the identity matrix, and ${\bf g}=[0,-g,0],$ with $g = 274.78\;\text{m}\cdot\text{s}^{-2}$ being the gravitational acceleration on the Sun.  Additionally, 
$\varrho_{\rm ie}\approx \varrho_{\rm i}$ and $\varrho_{\rm n}$ are 
the ion and neutral mass densities, 
$p_{\rm ie}=p_{\rm i}+p_{\rm e}=2p_{\rm i}$ and 
$p_{\rm n}$ are the gas pressures, 
$m_{\rm ie}\approx m_{\rm p}$, with $m_{\rm p}$ being the proton 
mass,  $m_{\rm n}$ represents the mass of each species, 
$T_{\rm ie}$ and $T_{\rm n}$ represent the temperatures, and $E_{\rm ie}$ and $E_{\rm n}$ are the total energy densities. Additionally, 
$k_{\rm B}$ is the Boltzmann constant, 
$\mu$ denotes the magnetic permeability, 
and $\gamma = 5/3$ is the adiabatic index. 
The symbol $\nu_{\rm in}$ represents 
the ion-neutral collision frequency, which is given as 
\citep{1965RvPP....1..205B, 2018SSRv..214...58B}
\begin{equation}
    \nu_{\rm in}=\frac{4}{3}\frac{\sigma_{\rm in}\varrho_{\rm n}}{m_{\rm ie}+m_{\rm n}}\sqrt{\frac{8k_{\rm B}}{\pi }\left(\frac{T_{\rm ie}}{m_{\rm ie}}+\frac{T_{\rm n}}{m_{\rm n}}\right)}\,.
\end{equation}
Here, $\sigma_{\rm in}$ represents the cross section of the ion-neutral collisions, with its magnitude $\sigma_{\rm in}=1.4\times 10^{-19}$~m$^{2}$ taken as its classical value from \citet{VranjesKrstic2013}. Moreover, $Q_{\rm i}$ and $Q_{\rm n}$ denote the heat production and exchange terms that result from ion-neutral collisions \citep{2018SSRv..214...58B}. The second terms on the right-hand side in Eqs.~(10) \& (11) describe the heat exchange between the ions and the neutrals.

The two-fluid equations consist of the conservation of mass (Eqs.~(1) $\&$ (2)), the momentum (Eqs.~(3) $\&$ (4)), and the energy (Eqs.~(5) $\&$ (6)) equations, which are completed by the induction equation and the solenoidal condition of Eq.~(7). In the given model, nonideal and nonadiabatic effects, ionization, recombination, radiation, viscosity, thermal conduction, and magnetic resistivity are not considered \citep{2019A&A...630A..79P, 2019ApJ...871....3S}. 
\subsection{Magnetohydrostatic equilibrium}
For computational economy it is assumed that the background solar atmosphere remains  at its magnetohydrostatic equilibrium ($\mathbf{V}_{\rm ie}=\mathbf{V}_{\rm n}={\bf 0}$). Then, from the momentum equations, it follows that 
    \begin{equation}
        -\nabla p_{\rm ie}+\varrho_{\rm ie}{\bf g} + \frac{1}{\mu}\left( \nabla \times \mathbf{B}\right)\times \mathbf{B}=\bf0\,,
    \end{equation}
    \begin{equation}
        -\nabla p_{\rm n}+\varrho_{\rm n}{\bf g}=\bf0\,.
    \end{equation}
The vertical profiles of the equilibrium gas pressures and mass densities at the magnetohydrostatic equilibrium state are given by \citep[e.g.,][]{2021MNRAS.506..989K}
\begin{eqnarray}
       && p_{\rm n}(y)=p_{\rm 0n} \exp \left(-\int_{y_{\rm r}}^{y}      \frac{dy}{\Lambda_{\rm n}(y)}\right)\,,\\
        && p_{\rm ie}(y)=p_{\rm 0ie} \exp \left(-\int_{y_{\rm r}}^{y}   \frac{dy}{\Lambda_{\rm i}(y)}\right)\,,\\
\hbox{and}&& \nonumber\\%
        &&\varrho_{\rm ie,n}(y)=\frac{p_{\rm ie,n}(y)}{g\Lambda_{\rm i,n}}\,,
        \hspace{10mm} \\
\hbox{with}&& \nonumber\\
       && \Lambda_{\rm n}=\frac{k_{\rm B} T(y)}{g m_{\rm n}}\,,
        \hspace{10mm}
        \Lambda_{\rm i}=\frac{k_{\rm B} T(y)}{g m_{\rm ie}}\,.
    \end{eqnarray}
Here, $\Lambda_{\rm n}$ and $\Lambda_{\rm i}$ denote the ion and neutral pressure scale heights, respectively. The symbols $p_{\rm 0n}=3\cdot 10^{-4}$~Pa and $p_{\rm 0ie}=10^{-2}$~Pa represent the neutral and charged gas pressures at the reference height $y_{\rm r}$, which is set at $y=50\;$Mm. 
The initial temperatures of ions and neutrals are set according to 
the semiempirical quiet solar atmosphere model of 
\citet{2008ApJS..175..229A}, that is $T_{\rm ie}(y)=T_{\rm n}(y)=T$ \citep{2016ApJ...818..128O}.

The gas pressures and mass densities profiles of Eqs.~(15)-(17) are overlaid by $\mathbf{B}=[0,B_{\rm y},B_{\rm z}]$. As a result of a nonzero value of the transversal component of magnetic field $B_{\rm z}$, the Alfv\'en and magnetoacoustic waves are linearly coupled \citep{2005LRSP....2....3N}. These waves decouple in the case of $B_{\rm z}=0$, in which the  $B_{\rm z}$ and $V_{\rm iz}$ perturbations correspond to Alfv\'en waves, while $V_{\rm ix}$ and $V_{\rm iy}$ perturbations are associated with magnetoacoustic waves.
\begin{figure}
   \begin{center}
        \vspace{0.2 cm}
        \includegraphics[width=0.4\textwidth]{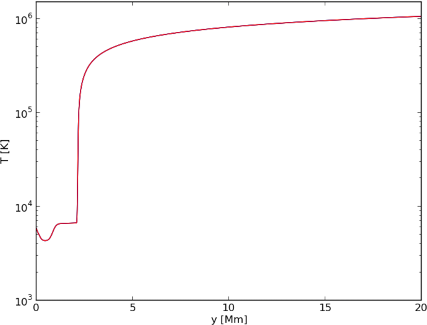}
        \vspace{0.2 cm}
        \includegraphics[width=0.4\textwidth]{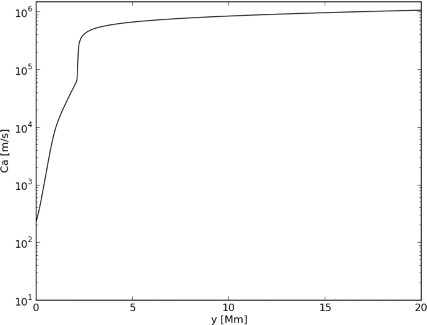}
        \hspace{0 cm}
        \includegraphics[width=0.4\textwidth]{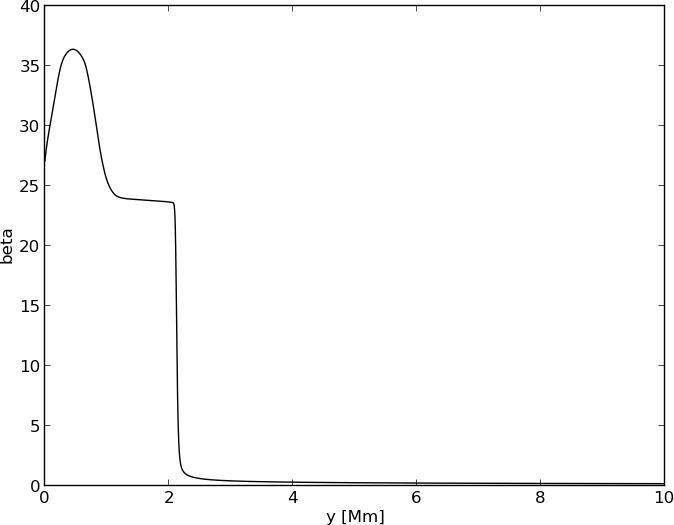}
    \end{center}
    \vspace{-0.5cm}
    \caption{Variation with height $y$ of the equilibrium temperature (top), Alfv\'en speed $c_{\rm a}$ (middle), and bulk plasma- $\beta$ (bottom) for $B_{\rm 0y} = 30\;$G and $B_{\rm 0z} = 5\;$G.}
    \label{fig:profiles}
\end{figure}

Figure~1 (top) shows the vertical profile of the initial equilibrium temperature $T$. It should be noted that this temperature reaches its minimum of $4341\;$K at $y=0.6\;$Mm, which is about 100 km above the bottom of the chromosphere. It rises to about $6000-7000\;$K in the middle and upper chromosphere ($1\;$Mm $\le y < 2.1\;$Mm). Then, in the transition region ($y \approx 2.1\;$Mm), the temperature rapidly increases and continues to rise with height in the solar corona until it reaches a magnitude of about $1\;$MK at $y=20\;$Mm. 

Figure~1 (middle) displays the bulk Alfv\'en speed, given by
    \begin{equation}
        c_{\rm a}=\frac{|B|}{\sqrt{\mu\varrho_{\rm i}}}\,.
    \end{equation}
It grows with altitude, and its sudden increase (from about $60\; \text{to}\; 400\;\text{km}\cdot\text{s}^{-1}$) occurs in the transition region. Still, in contrast to the temperature profile, there is no decrease in$c_{\rm a}$ in the middle of the photosphere, and the minimum of $c_{\rm a}$ of about $200-300\;\text{m}\cdot\text{s}^{-1}$ takes place at the bottom of the photosphere ($y = 0\;$Mm). In the corona, $c_{\rm a}$ slowly and continuously grows until at $y=20\;$Mm, $c_{\rm a}$ attains its value of about $10^{3}\;\text{km}\cdot\text{s}^{-1}$. 

Figure~1 (bottom) illustrates the bulk plasma-$\beta$, which is the ratio of ion
 + electron and neutral thermal pressures to magnetic pressure:
    \begin{equation}
        \beta=\frac{p_{\rm i e}+p_{\rm n}}{B^2/2\mu_{\rm 0}}.
    \end{equation}
The given plot demonstrates that the plasma-$\beta$ trend is reversed to that of the temperature. At the height $(y\approx 0.6\;\text{Mm})$ that corresponds to the temperature minimum, the plasma-$\beta$ attains its local maximum of about $37$. In the chromosphere, the plasma-$\beta$ falls off to about $24$, while there is a sudden decrease in its magnitude in the transition region. In the solar corona, the plasma-$\beta$ experiences an abrupt drop with height to $\beta<1$. 
\section{Numerical simulations}
Aiming to study two-fluid linearly coupled Alfv\'en and magnetoacoustic waves in a gravitationally stratified and partially ionized photosphere and chromosphere, numerical simulations were performed with the use of the JOANNA code \citep{2018MNRAS.481..262W, 2019ApJ...884..127W}. This code solves the initial-boundary value problem for the two-fluid equations numerically in the form of Eqs.~(1) - (11). In the simulations, the Courant-Friedrichs-Lewy number \citep{1928MatAn.100...32C} was set to 0.9. The second-order accurate linear spatial reconstruction \citep{2009JCoPh.228.3368T} and the third-order accurate Super Stability Preserving Runge-Kutta (SSPRK3) method \citep{2010nmfd.book.....D} were used. This was extended by applying the Harten-Lax-van Leer Discontinuity (HLLD) approximate Riemann solver \citep{2005AGUFMSM51B1295M}. Besides, the divergence of the magnetic field cleaning method of \citet{2002JCoPh.175..645D} was implied.
\subsection{Numerical box and boundary conditions}
The two-dimensional simulation domain was defined as $-0.08\; \text{Mm} \le x \le 0.08\;$Mm along the horizontal $x$-direction and $-0.5\; \text{Mm} \le y \le 60\;$Mm along the vertical $y$-direction. The whole box in the $x$-direction was covered by 16 cells, with cell size $\Delta x = 10\;$km. The region $-0.5\; \text{Mm} \le y \le 4.62\;$Mm was covered by a uniform grid of 2048 cells, so vertical cell size $\Delta y = 2.5\;$km. However, the upper zone of the simulation box, specified by $4.62\; \text{Mm} \le y \le 60\;$Mm, was divided into 32 cells of the nonuniform grid. Here, the size of the cells steadily grows with height, so the grid was stretched along the $y$-direction. This stretched grid damped any incoming signal from the top boundary, reducing inherent reflections from the level of $y=60\;$Mm \citep[e.g.,][]{2018ApJ...866...50K}. At this level and at the bottom of the simulation box, all plasma variables were set equal to their magnetohydrostatic equilibrium values. Along the ($x-$) boundaries, ``open'' boundary conditions were implemented, which means that the $x$-derivatives of all the plasma quantities were set equal to zero at the left- and right-hand sides of the simulation domain.
\subsection{Impulsive perturbations}
Intending to perturb the magnetohydrostatic equilibrium, a pulse in the transverse components of ion and neutral velocities, $V_{\rm iz}$ and $V_{\rm nz}$, was launched initially (at $t = 0\;$s):   
    \begin{equation}
      V_{\rm iz}(x,y, t=0\;\text{s})=V_{\rm nz}(x,y, t=0\;\text{s})=A\; \exp\left(-\frac{\mathbf{x^2}+(y-y_{\rm 0})^2}{w^2}\right)\,.
    \end{equation}
\begin{figure*}
    \begin{center}
        \hspace{0cm}
        \includegraphics[width=0.4\textwidth]{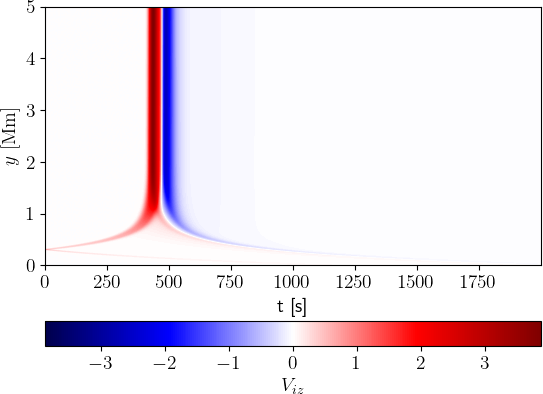}
        \hspace{1cm}
        \includegraphics[width=0.4\textwidth]{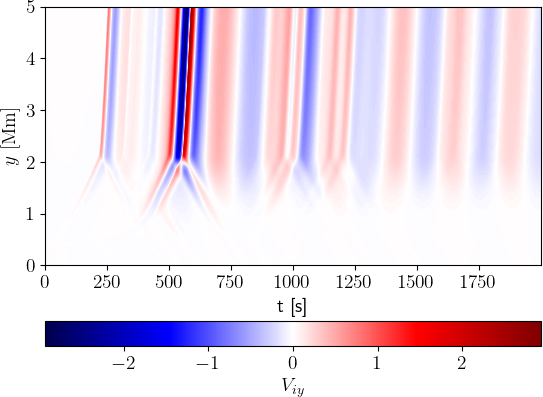}
        \hspace{1cm}
        \includegraphics[width=0.4\textwidth]{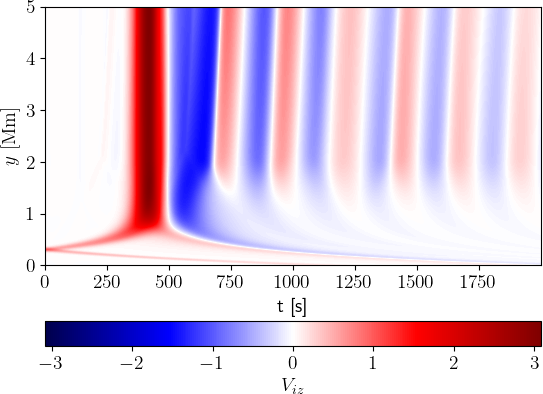}
        \hspace{1cm}
        \includegraphics[width=0.4\textwidth]{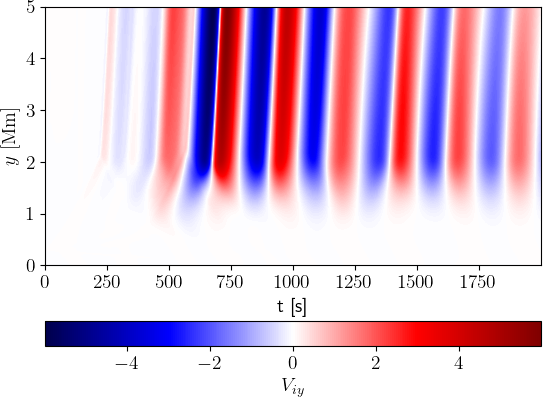}
        \vspace{1cm}
        \includegraphics[width=0.4\textwidth]{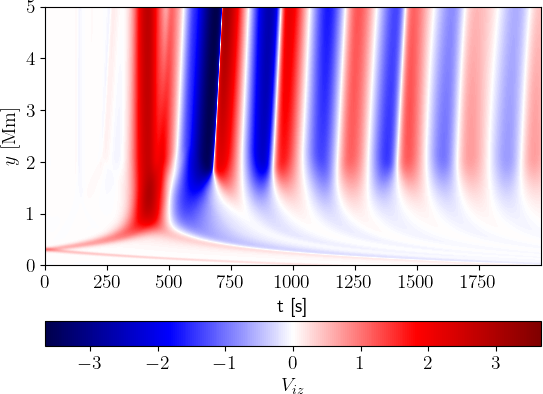}
        \hspace{1cm}
        \includegraphics[width=0.4\textwidth]{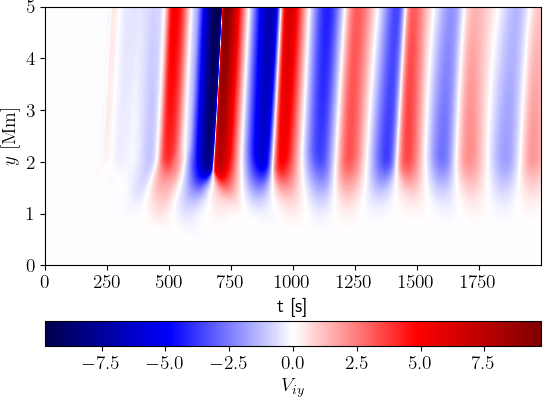}
        \vspace{-1.5cm}
    \end{center}
    \vspace{-0cm}
    \caption{Time–distance plots for $V_{\rm iz}$ (left) and $V_{\rm iy}$ (right),  in the case of $B_{\rm 0z} = 0\;$G (top), $B_{\rm 0z} = 5\;$G (middle), and $B_{\rm 0z} = 10\;$G (bottom) for $B_{\rm 0y} = 30\;$G, $y_{\rm 0} = 0.3\;$Mm, $A=1\;\text{km}\cdot\text{s}^{-1}$, and  $w=0.1\;$Mm.}
\end{figure*} 
Here, $A$ represents the amplitude of the pulse, $w$ is its width, and $y_{\rm 0}$ denotes the pulse's location along $y$. Based on Paper~I, in most of the considered simulations, the localization of the initial pulse was chosen at $y_{\rm 0}=0.3\;$Mm. The values of $A$ and $w$ varied in the simulations. In this paper, $A=1\;\text{km}\cdot\text{s}^{-1}$ and $A=10\;\text{km}\cdot\text{s}^{-1}$ were chosen and the pulse width varied from $w=0.05\;$Mm to $w=0.2\;$Mm.

\subsection{Small-amplitude case results}
This subsection looks at small-amplitude two-fluid Alfv\'en and magnetoacoustic waves. Figure~2 shows the evolution of these waves, which are excited at the height $y_{\rm 0} = 0.3\;$Mm, by the initial pulse with an amplitude equal to $A=1\;$km$\cdot$s$^{-1}$ and with a width $w=0.1\;$Mm. In this case, the vertical magnetic field is fixed and equal to $B_{\rm 0y}=30\;$G, and the transverse magnetic field $B_{\rm 0z}$ varies from $0\;$G (top), through $5\;$G (middle), to $10\;$G (bottom).

The left panels of Fig.~2 reveal that the initial pulse splits into two counter-propagating waves that are damped by ion-neutral collisions. Also, the upwardly propagating waves experience partial reflection at about $t=500\;$s, which takes place at the height $y \approx 0.8\;$Mm, corresponding to the low chromosphere. These facts mean that the upwardly propagating waves travel with similar speeds. Simple calculations show that the signals' propagation speed is about $1\;$km$\cdot$s$^{-1}$, which agrees well with Fig.~1 (middle), where the average Alfv\'en speed $c_{\rm a}$ is about $1\;$km$\cdot$s$^{-1}$ at $y \approx 0.8\;$Mm. 
\begin{figure*}[h]
    \begin{center}
        \hspace{0cm}
        \includegraphics[width=0.4\textwidth]{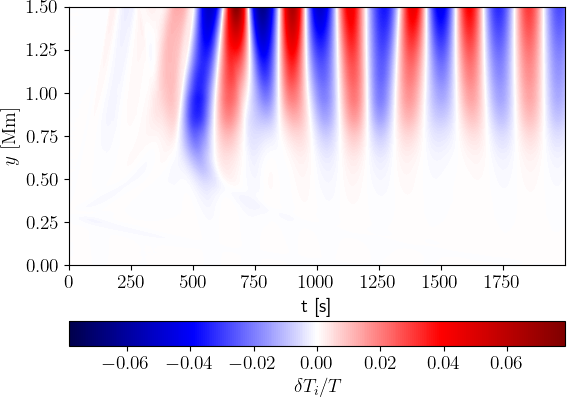}
        \hspace{1cm}
        \includegraphics[width=0.38\textwidth]{Viy_by30-bz5_y03_a1_w01.png}
        \hspace{1cm}
        \includegraphics[width=0.38\textwidth]{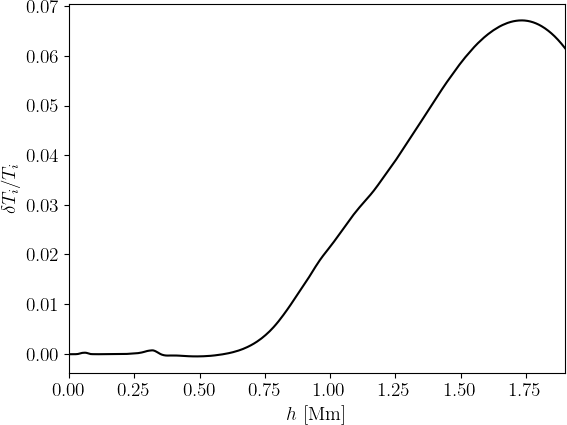}
        \hspace{1cm}
        \includegraphics[width=0.38\textwidth]{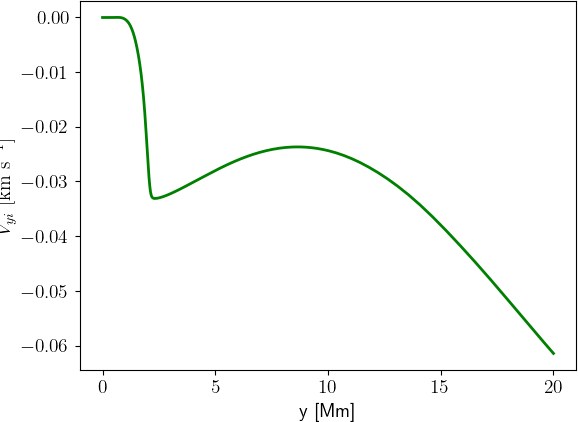}
        \vspace{-0.5cm}
    \end{center}
    \vspace{-0cm}
    \caption{Time-distance plots for $\delta T_{ie}/T$ (top left), and $V_{\rm iy}$ (top right) and vertical profiles of $<\delta T_{ie}/T>_{\rm t}$ (bottom left) and $<V_{\rm iy}>_{\rm t}$ (bottom right), both averaged over time, in the case of $B_{\rm 0y} = 30$, $B_{\rm 0z} = 5\;$G, $y_{\rm 0} = 0.3\;$Mm,  $w=0.1\;$Mm, and $A=1\;\text{km}\cdot\text{s}^{-1}$.}
\end{figure*}

In these panels it is also demonstrated that max$(V_{\rm iz}) \approx 4\;$km$\cdot$s$^{-1}$ and it is almost equal in the cases of $B_{\rm 0z}=0\;$G and $B_{\rm 0z}=10\;$G. However, this value is slightly smaller for $B_{\rm 0z}=5\;$G, with max$(V_{\rm iz}) \approx 3\;$km$\cdot$s$^{-1}$. These results can be compared to the left panels of Fig.~3 in \citet{2021A&A...652A.114P}, which correspond to $B_{\rm 0z}=0\;$G, $B_{\rm 0y}=30\;$G, $A=1\;$km$\cdot$s$^{-1}$, $w=0.2\;$Mm, $y_{\rm 0} = 0\;$Mm, and $y_{\rm 0} = 0.5\;$Mm. From there, it is clear that max$(V_{\rm iz}$) is in the range of $2.5 - 3.4\;$km$\cdot$s$^{-1}$. The current results are slightly higher, even with a pulse width that is  two times smaller than the older results.

The right panels of Fig.~2 show the vertical component of the ion velocity, $V_{\rm iy}$, versus time. It is clearly seen that the maximum value of the vertical velocity component max$(V_{\rm iy})$ grows with the magnitude of the transversal magnetic field value $B_{\rm 0z}$. In the case of $B_{\rm 0z}=0\;$G, max$(V_{\rm iy}) \approx 3\;$km$\cdot$s$^{-1}$, then for $B_{\rm 0z}=5\;$G,  max$(V_{\rm iy})=6\;$km$\cdot$s$^{-1}$, and lastly for $B_{\rm 0z}=10\;$G, max$(V_{\rm iy})=10\;$km$\cdot$s$^{-1}$. 
From these plots, it can be noted that the initial pulse also splits into two counter-propagating waves and is damped by ion-neutral collisions, as it is well seen in the adjacent panels. This splitting occurs at almost the same time ($t \approx 250\;$s) and at the same height ($y \approx 1.9\;$Mm) for different transversal magnetic field values. From this, the wave propagation speed can be estimated, and its value is about $7.6\;$km$\cdot$s$^{-1}$.

Figure~3 (top panels) illustrates time-distance plots for the relative perturbed ion temperature $\delta T_{\rm ie}/T$ (left panel) and for the vertical component of the ion velocity $V_{\rm iy}$ (right panel) in the cases of $B_{\rm 0y} = 30\;$G, $B_{\rm 0z} = 5\;$G, $y_{\rm 0} = 0.3\;$Mm,  $A=1\;\text{km}\cdot\text{s}^{-1}$, and $w=0.1\;$Mm. There are strong correlations between the velocity and temperature signals in the small range of $y$, about $0 - 1.5\;$Mm (the plots look almost identical). The top left panel reveals that the maximum value of the perturbed relative ion temperature $\max(\delta T_{\rm ie}/T)$, is about $0.08\;$K, whereas the top right plot shows that the maximum value of the vertical component of ion velocity $\max(V_{\rm iy})$ is about $6\;$km$\cdot$s$^{-1}$. The result presented in the top left plot can be compared with a similar case from Paper~I. The only difference between the cases is the value of the transversal magnetic field (that is, the presence of magnetoacoustic waves in the present simulations). As it was already said in the case of a transversal magnetic field  $B_{\rm 0z}=5\;$G -- $\max(\delta T_{\rm ie}/T) \approx 0.08\;$K, in the case of $B_{\rm 0z}=0\;$G (from Paper~I), the result is slightly smaller ($\max(\delta T_{\rm ie}/T) \approx 0.05\;$K). On the other hand, the maximum value of the vertical ion velocity component is smaller for zero transversal magnetic field, so there $\max(V_{\rm iy})\approx 3.6 \;$km$\cdot$s$^{-1}$, and in the current case it is about $\max(V_{\rm iy})=6 \;$km$\cdot$s$^{-1}$.
\begin{figure*}
    \begin{center}
        \hspace{-0cm}
        \includegraphics[width=0.38\textwidth]{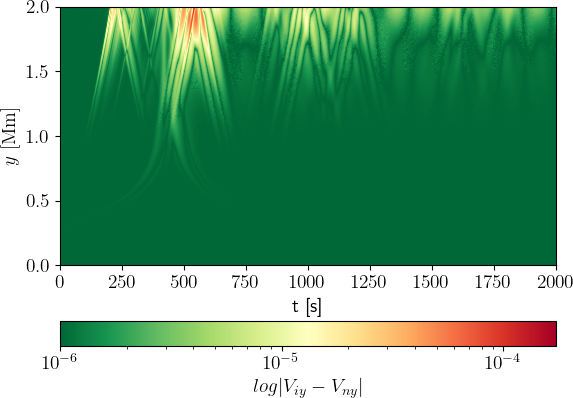}
        \hspace{1cm}
        \includegraphics[width=0.38\textwidth]{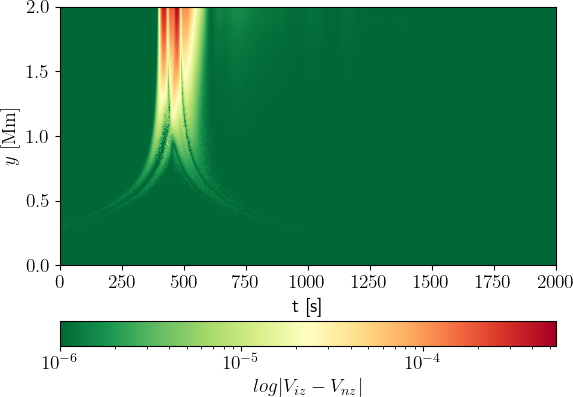}
        \hspace{1cm} 
        \includegraphics[width=0.38\textwidth]{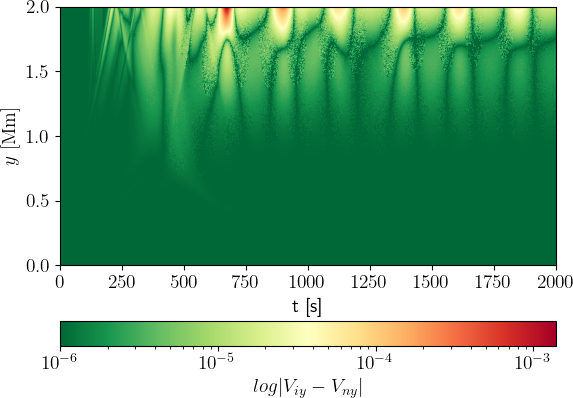}
        \hspace{1cm}
        \includegraphics[width=0.38\textwidth]{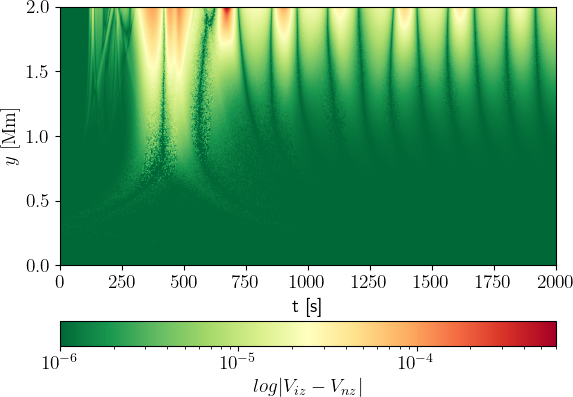}
        \hspace{1cm}
        \includegraphics[width=0.38\textwidth]{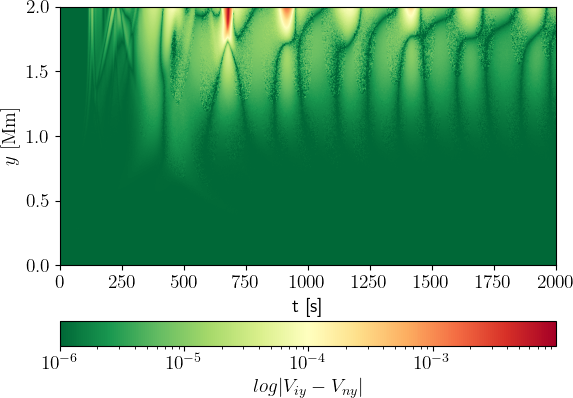}
        \hspace{1cm}
        \includegraphics[width=0.38\textwidth]{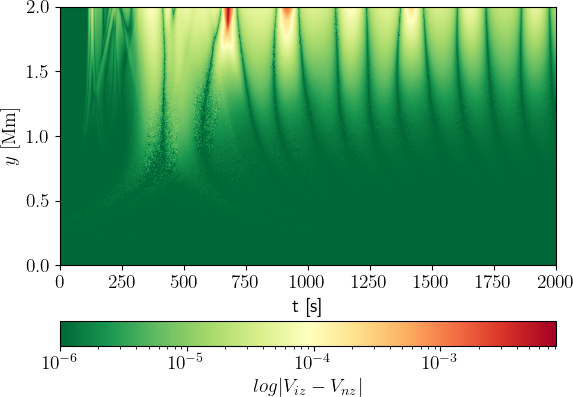}
        \vspace{0cm}
    \end{center}
    \vspace{-0.5cm}
    \caption{Velocity drifts for $V_{\rm iy} - V_{\rm ny}$ (left) and $V_{\rm iz} - V_{\rm nz}$ (right), in the cases of $B_{\rm 0z} = 0\;$G (top), $B_{\rm 0z} = 5\;$G (middle), and $B_{\rm 0z} = 10\;$G (bottom) for $B_{\rm 0y} = 30\;$G, $A=1\;\text{km}\cdot\text{s}^{-1}$, $w=0.1\;$Mm, and $y_{\rm 0} = 0.3\;$Mm. The velocity drifts are expressed in units of $1\;\text{km}\cdot\text{s}^{-1}$.}
\end{figure*} 

The bottom panels of Fig.~3 demonstrate the temporarily averaged relative perturbed temperature $\left\langle  {\delta T_{ie}}/{T} \right\rangle_{\rm t}$ and vertical ion velocity $\langle V_{\rm iy}\rangle_{\rm t}$, which can be defined as
\begin{equation}
    \left\langle  \frac{\delta T_{ie}}{T} \right\rangle_{\rm t}  = \frac{1}{t_{\rm 2} - t_{\rm 1}} \int_{t_{\rm 1}}^{t_{\rm 2}} \frac{\delta T_i-e}{T} \; dt \, ,
\end{equation}
\begin{equation}
    \langle V_{\rm iy}\rangle_{\rm t}= \frac{1}{t_{\rm 2} - t_{\rm 1}} \int_{t_{\rm 1}}^{t_{\rm 2}} V_{\rm iy} \; dt \, ,
\end{equation}
where $t_{\rm 1}=0\;$s and $t_{\rm 2}=3000\;$s. Because of the small amplitude of the initial pulse ($A=1\;\text{km}\cdot\text{s}^{-1}$), the values for low $y$ are negligibly small. In the case of the bottom left panel, $\left\langle  {\delta T_{ie}}/{T} \right\rangle_{\rm t}$ reaches a maximum of only about $0.07\;$K in the lower corona (but higher up it starts to decrease). Also, a small bump is noticeable at $y \approx 0.3\;$Mm (this is the altitude where the initial pulse is launched from). The bottom right panel shows that up to $y \approx 20\;$Mm, a down-flow occurs with its minimum velocity of about $-0.06\;$km$\cdot$s$^{-1}$.

Figure~4 presents the ion-neutral velocity drifts for the vertical velocity components, $V_{\rm iy}-V_{\rm ny}$, in the left panels and transversal components, $V_{\rm iz}-V_{\rm nz}$, in the right panel, for $B_{\rm 0y} = 30\;$G, $A=1\;\text{km}\cdot\text{s}^{-1}$, $w=0.1\;$Mm, and $y_{\rm 0} = 0.3\;$Mm. These plots differ due to the transversal magnetic field value, which varies from $B_{\rm 0z}=0\;$G (top), via $B_{\rm 0z}=5\;$G (middle), to $B_{\rm 0z}=10\;$G (bottom). It is clearly seen that maximum values of velocity drift grow with $B_{\rm 0z}$. For $V_{\rm iy}-V_{\rm ny}$ (left panels), its maximum value increases from about $10^{-4}$ to $10^{-2}\;\text{km}\cdot\text{s}^{-1}$, and for $V_{\rm iz}-V_{\rm nz}$ (right panels) from about $10^{-4}$ to almost $10^{-2}\;\text{km}\cdot\text{s}^{-1}$. In every case, it can be noticed that the velocity drifts grow with height. Additionally, for lower heights (in the range $0-0.5\;$Mm), the velocity drift values are very small, indicating that $V_{\rm iz}$ almost equals $V_{\rm nz}$, and the same for $V_{\rm iy}$ and $V_{\rm ny}$. This is due to the fact that ions and neutrals are strongly coupled in the lower atmosphere. The highest velocity drift values are achieved in the time range $0 - 1000\;$s; hence, it follows that plasma heating occurs in the initial phase after the waves are generated. Another point is that the maximum values of velocity drift are greater for $V_{\rm iz}-V_{\rm nz}$ than for $V_{\rm iy}-V_{\rm ny}$ for $B_{\rm 0z} = 0\;$G, and they are equal or even smaller for $B_{\rm 0z} = 5\;$G and $B_{\rm 0z} = 10\;$G. Due to this fact, it can be stated that Alfv\'en waves are responsible for the plasma heating for the small amplitude and transversal magnetic field-free case.
\begin{figure}
    \begin{center}
       \hspace{-1cm}
      \includegraphics[width=0.48\textwidth]{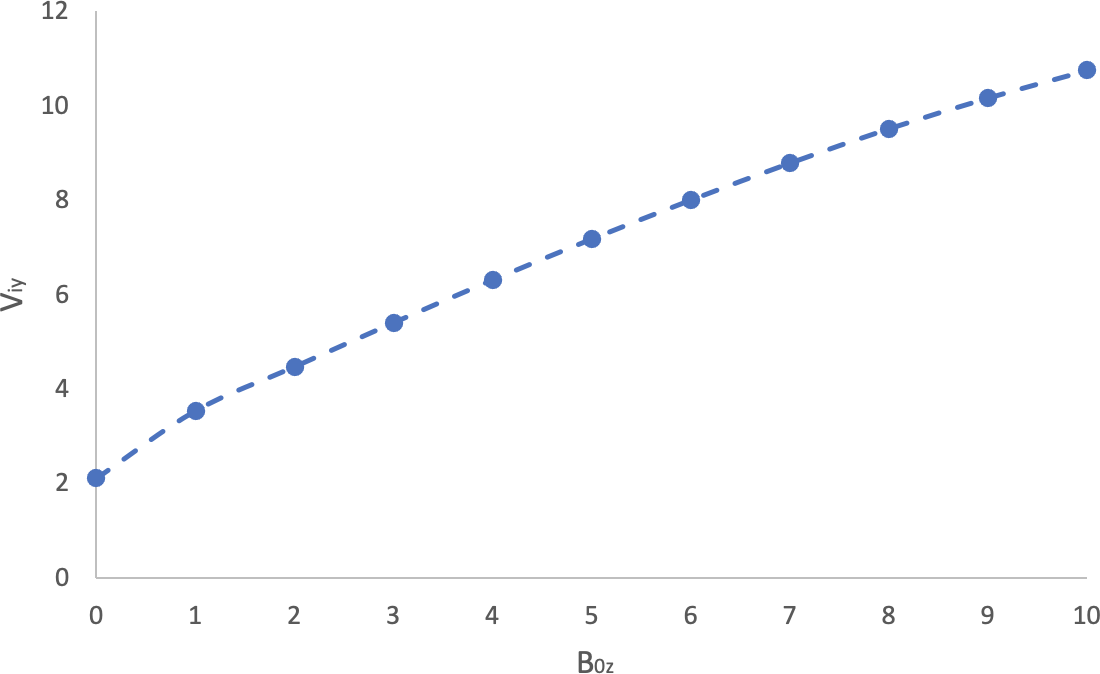}
        \vspace{0cm}
    \end{center}
    \vspace{-0.5cm}
    \caption{Maximum of the vertical component of the ion velocity, $V_{\rm iy}$, vs. the transverse magnetic field $B_{\rm 0z}$, for $B_{\rm 0y} = 30\;$G, $y_{\rm 0}=0.3\;$Mm, $A = 1\;\text{km}\cdot\text{s}^{-1}$, and $w=0.1\;$Mm.}
\end{figure} 
\begin{figure*}
    \begin{center}
        \hspace{0cm}
        \includegraphics[width=0.4\textwidth]{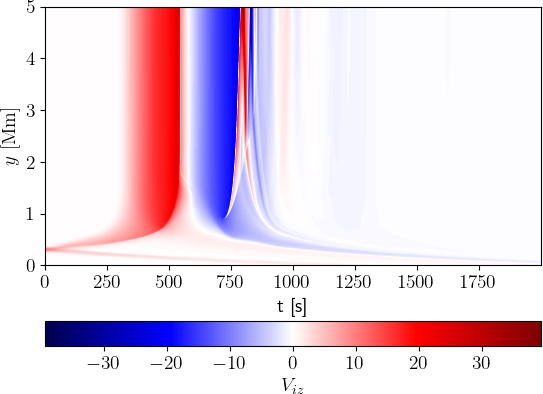}
        \hspace{1cm}
        \includegraphics[width=0.4\textwidth]{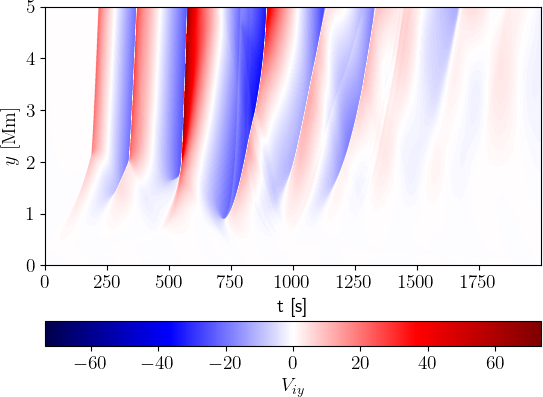}
        \hspace{1cm}
        \includegraphics[width=0.4\textwidth]{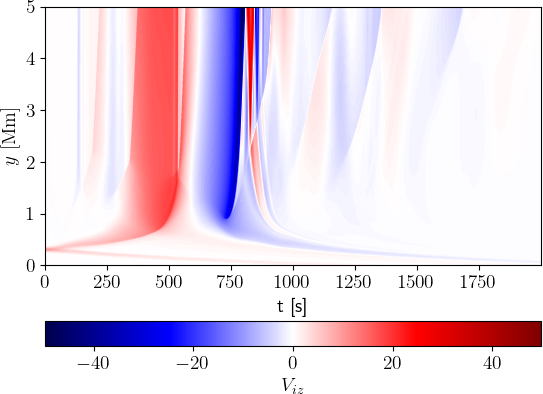}
        \hspace{1cm}
        \includegraphics[width=0.4\textwidth]{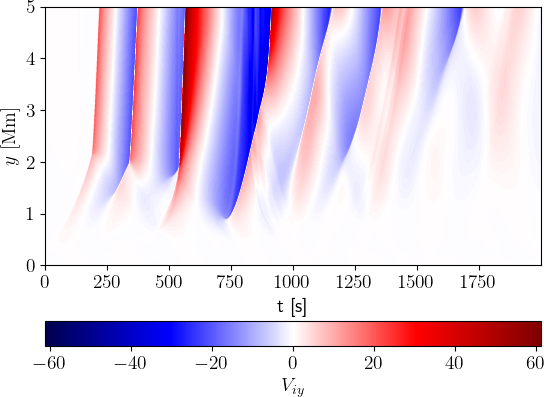}
        \hspace{1cm}
        \includegraphics[width=0.4\textwidth]{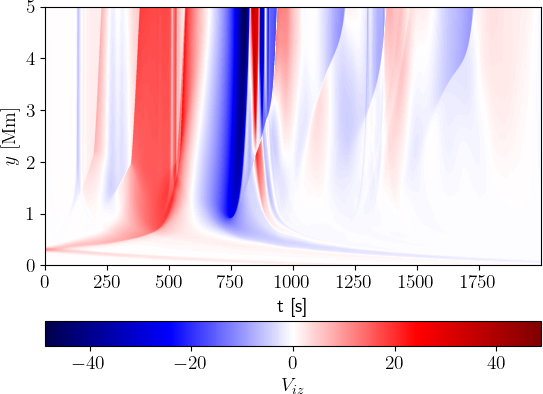}
        \hspace{1cm}
        \includegraphics[width=0.4\textwidth]{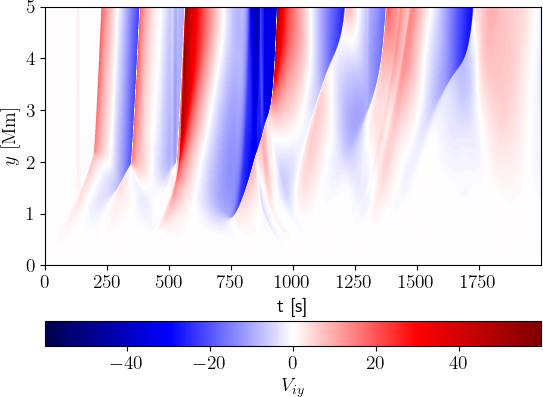}
        \vspace{-0.5cm}
    \end{center}
    \vspace{-0cm}
    \caption{Time–distance plots for $V_{\rm iz}$ (left) and $V_{\rm iy}$ (right) in the case of $B_{\rm 0z} = 0\;$G (top), $B_{\rm 0z} = 5\;$G (middle), and $B_{\rm 0z} = 10\;$G (bottom) for $B_{\rm 0y} = 30\;$G, $A=10\;\text{km}\cdot\text{s}^{-1}$, $w=0.2\;$Mm, and $y_{\rm 0} = 0.3\;$Mm.}
\end{figure*} 

Figure~5 illustrates the variation of the max value of the vertical component of the ion velocity $V_{\rm iy}$ for different values of the transverse magnetic field $B_{\rm 0z}$ for $B_{\rm 0y} = 30\;$G, $y_{\rm 0}=0.3\;$Mm, $A = 1\;\text{km}\cdot\text{s}^{-1}$, and $w=0.1\;$Mm. From this plot it can be inferred that $\max(V_{\rm iy})$ is directly dependent on $B_{\rm 0z}$ (i.e.\ $\max(V_{\rm iy})$ grows with $B_{\rm 0z}$). This growth can be compared to the outcome of Fig.~2. For instance, from Fig.~2 (middle-right panel) for $B_{\rm 0z}=5\;$G, max$(V_{\rm iy})$ is about $5\;$km$\cdot$s$^{-1}$, and from the current plot its value is about $7\;$km$\cdot$s$^{-1}$. The present values are slightly larger and this results from a larger $y$ range (here it is up to $20\;$Mm).
\subsection{Large-amplitude case results}
\begin{figure*}
    \begin{center}
        \hspace{-0cm}
        \includegraphics[width=0.4\textwidth]{Viz_by30-bz0_y03_a10_w02.png}
        \hspace{1cm}
        \includegraphics[width=0.4\textwidth]{Viy_by30-bz0_y03_a10_w02.png}
        \hspace{1cm}
        \includegraphics[width=0.4\textwidth]{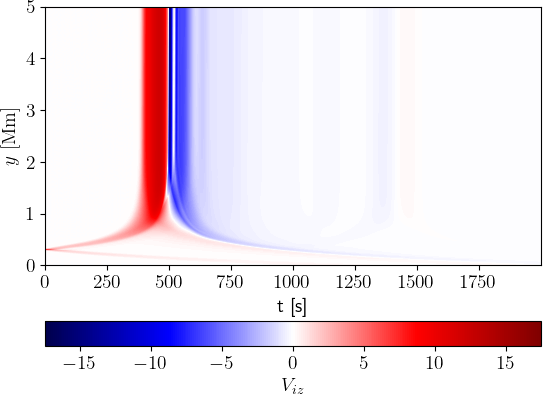}
        \hspace{1cm}
        \includegraphics[width=0.4\textwidth]{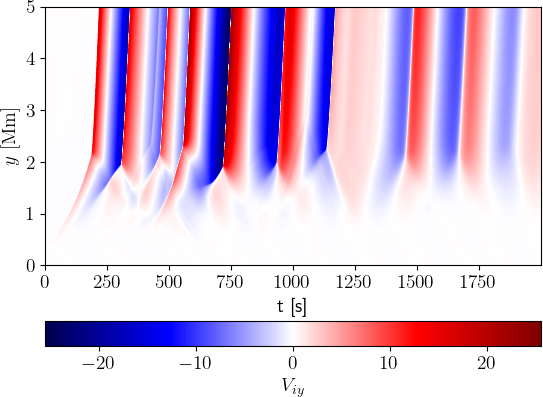}
        \hspace{1cm}
        \includegraphics[width=0.4\textwidth]{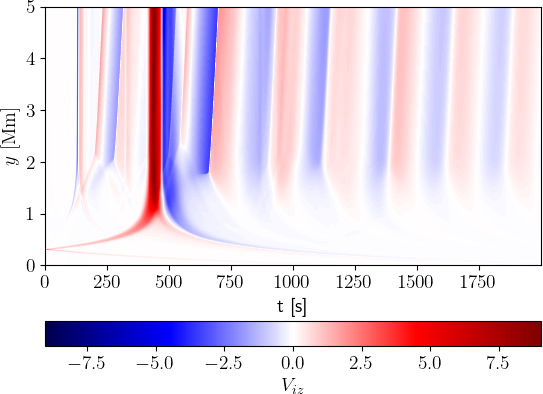}
        \hspace{1cm}
        \includegraphics[width=0.4\textwidth]{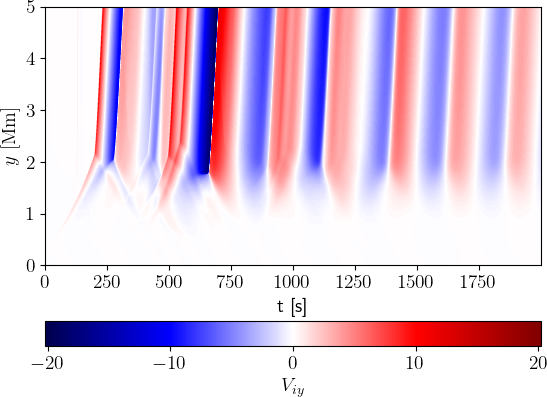}
        \vspace{-0.5cm}
    \end{center}
    \vspace{-0cm}
    \caption{Time–distance plots for $V_{\rm iz}$ (left) and $V_{\rm iy}$ (right), in the case of $w=0.2\;$Mm (top), and $w=0.1\;$Mm (middle), $w=0.05\;$Mm (bottom) for $B_{\rm 0y} = 30\;$G, $B_{\rm 0z} = 5\;$G, $A=10\;\text{km}\cdot\text{s}^{-1}$, and $y_{\rm 0} = 0.3\;$Mm.}
\end{figure*} 
This part of the paper presents results for the case of a much larger amplitude pulse than in the earlier subsection. Figure~6 shows the evolution of Alfv\'en and magnetoacoustic waves that are excited by the initial pulse of its amplitude $A=10\;\text{km}\cdot\text{s}^{-1}$, and launched from the photosphere at $y_{\rm 0}=0.3\;$Mm. The vertical magnetic field is equal to $B_{\rm 0y}=30\;$G and the transversal magnetic field varies from $B_{\rm 0z}$ being $0\;$G (top), through $5\;$G (middle), to $10\;$G (bottom). A noticeable difference in the transversal (left panels) and vertical (right panels) ion velocity components is the velocity maximum. For $B_{\rm 0z}=0\;$G, max$(V_{\rm iz})\approx40\;\text{km}\cdot\text{s}^{-1}$, for $B_{\rm 0z}=5\;$G, it attains its highest value at about $50\;\text{km}\cdot\text{s}^{-1}$, and in the case of $B_{\rm 0z}=10\;$G, it falls off to $\approx45\;\text{km}\cdot\text{s}^{-1}$. These outcomes can be compared to the results of \citet{2021A&A...652A.114P}, where only Alfv\'en waves were considered for $B_{\rm 0y}=30\;$G, $B_{\rm 0z}=0\;$G, $A=10\;\text{km}\cdot\text{s}^{-1}$, $w=0.1\;$Mm, and $y_{\rm 0} = 0.3\;$Mm (Fig.~4 top left panel); there, max$(V_{\rm iz}) \approx 27\;\text{km}\cdot\text{s}^{-1}$. The change in the max$(V_{\rm iz})$ value is due to the larger pulse width in the current simulations. 

It follows from the right panels of Fig.~6 that max$(V_{\rm iy})$ reaches the highest value of about $73\;\text{km}\cdot\text{s}^{-1}$ in the case of $B_{\rm 0z}=0\;$G, then it slightly decreases to $\approx 62\;\text{km}\cdot\text{s}^{-1}$ for $B_{\rm 0z}=5\;$G, and it attains a value of about $60\;\text{km}\cdot\text{s}^{-1}$ for $B_{\rm 0z}=10\;$G. These results can also be compared to the almost identical case from \citet{2021A&A...652A.114P}. We notice that, despite the pulse width being twice as large in the present case, the maximum value of the vertical ion velocity component remains the same: max$(V_{\rm iy})\approx 70\;\text{km}\cdot\text{s}^{-1}$ (Fig.~5, top right panel).

Figure~7 displays transverse (left panels) and vertical (right panels) ion velocity components for $B_{\rm 0y} = 30\;$G, $B_{\rm 0z} = 5\;$G, $A=10\;\text{km}\cdot\text{s}^{-1}$, and $y_{\rm 0} = 0.3\;$Mm. In this figure the varying value is the width of the initial pulse, and mainly it changes from $w=0.2\;$Mm (top), over $w=0.1\;$Mm (middle), to $w=0.05\;$Mm (bottom). In both the left and right panels, it is apparent that the maximum values of these velocities grow with $w$. For instance, max$(V_{\rm iz})$ is in the range of about $7.7-40\;$km$\cdot$s$^{-1}$, and max$(V_{\rm iy})$ changes in the range of about $20-72\;$km$\cdot$s$^{-1}$. We note that the height in which the $V_{\rm iz}$ and $V_{\rm iy}$ signals are partially reflected remains independent of $w$, but the time changes slightly (it is difficult to obtain precise values from the plots). From this fact, it follows that the waves' speed rises with $w$.
\begin{figure*}
    \begin{center}
        \hspace{-0cm}
        \includegraphics[width=0.4\textwidth]{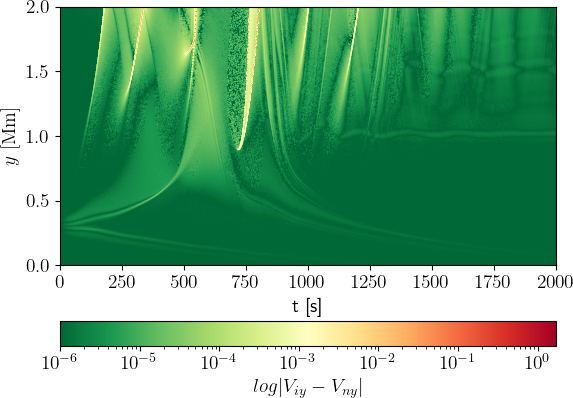}
        \hspace{1cm}
        \includegraphics[width=0.4\textwidth]{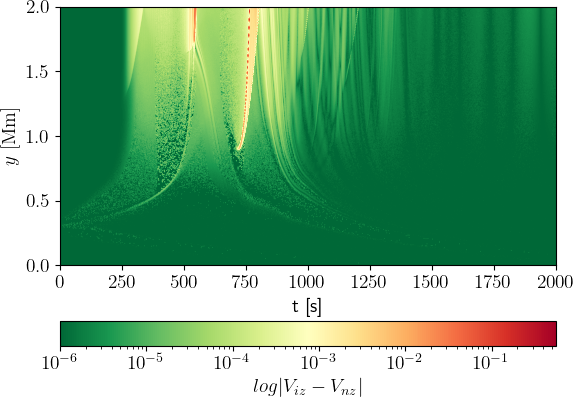}
        \hspace{1cm}
        \includegraphics[width=0.4\textwidth]{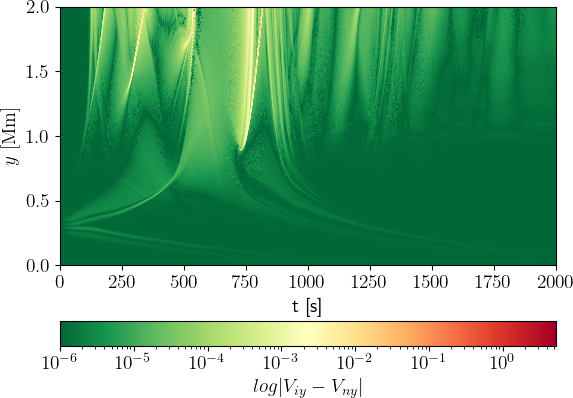}
        \hspace{1cm}
        \includegraphics[width=0.4\textwidth]{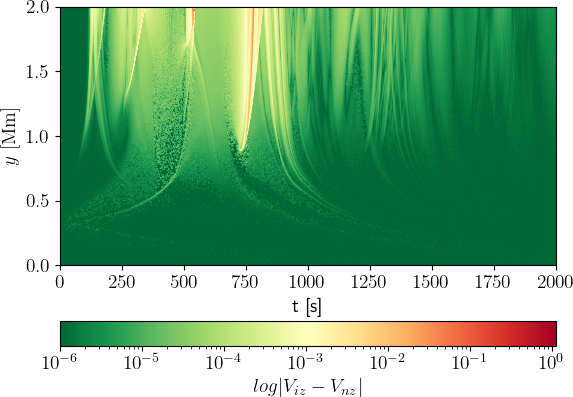}
        \hspace{1cm}
        \includegraphics[width=0.4\textwidth]{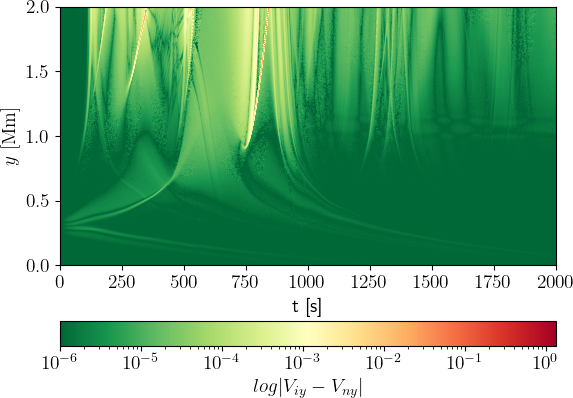}
        \hspace{1cm}
        \includegraphics[width=0.4\textwidth]{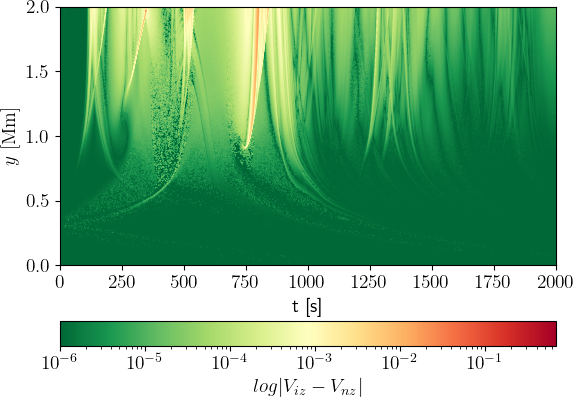}
        \vspace{-0.5cm}
    \end{center}
    \vspace{-0cm}
    \caption{Velocity drifts for $V_{\rm iy} - V_{\rm ny}$ (left) and $V_{\rm iz} - V_{\rm nz}$ (right), in the case of $B_{\rm 0z} = 0\;$G (top), $B_{\rm 0z} = 5\;$G (middle), and $B_{\rm 0z} = 10\;$G (bottom) for $B_{\rm 0y} = 30\;$G, $A=10\;\text{km}\cdot\text{s}^{-1}$, $w=0.2\;$Mm, and $y_{\rm 0} = 0.3\;$Mm.}
\end{figure*} 
\begin{figure*}
    \begin{center}
        \hspace{0cm}
        \includegraphics[width=0.37\textwidth]{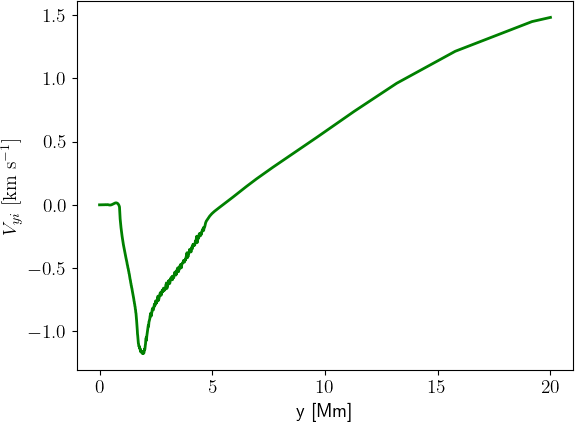}
        \hspace{0cm}
        \includegraphics[width=0.37\textwidth]{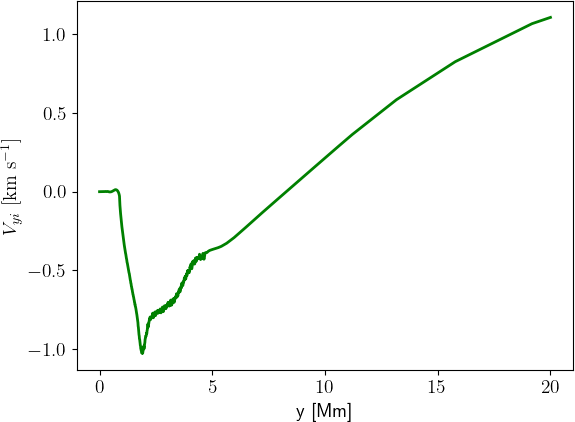}
        \hspace{0cm}
        \includegraphics[width=0.37\textwidth]{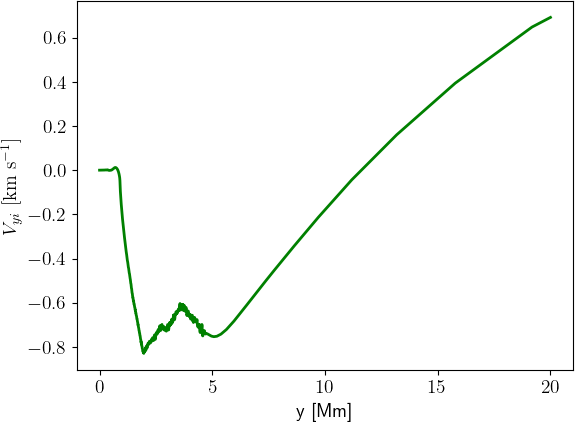}
    \end{center}
    \vspace{-0.5cm}
    \caption{Time–distance plots for averaged in time $V_{\rm iy}$ in the case of $B_{\rm 0z} = 0\;$G (top left), $B_{\rm 0z} = 5\;$G (top right), and $B_{\rm 0z} = 10\;$G (bottom) for $B_{\rm 0y} = 30\;$G, $A=10\;\text{km}\cdot\text{s}^{-1}$, and $y_{\rm 0} = 0.3\;$Mm.}
\end{figure*} 

Figure~8 illustrates the ion-neutral velocity drifts for the vertical velocity component, $V_{\rm iy}-V_{\rm ny}$ (left panels) and the transverse component, $V_{\rm iz}-V_{\rm nz}$ (right panels). This corresponds to $B_{\rm 0y} = 30\;$G, $A=10\;\text{km}\cdot\text{s}^{-1}$, $w=0.2\;$Mm, and $y_{\rm 0} = 0.3\;$Mm, with a varying magnitude of the transversal magnetic field $B_{\rm 0z}$ (in the range $0-10\;$G). It is clearly seen that the maximum value of velocity drift remains equal for almost every simulation and it is about $1\;\text{km}\cdot\text{s}^{-1}$, except for $V_{\rm iy}-V_{\rm ny}$ in the case of $B_{\rm 0z} = 5\;$G (middle-left panel), where the maximum value reaches $15\;\text{km}\cdot\text{s}^{-1}$. Besides, in this particular case, $V_{\rm iy}-V_{\rm ny}$ is greater than $V_{\rm iz}-V_{\rm nz}$, and due to this fact, it can be noted that magnetoacoustic waves are responsible for the plasma heating in the case of a large amplitude and a transversal magnetic field of $5\;$G.

Figure~9 demonstrates the temporarily averaged vertical ion velocity $\langle V_{\rm iy}\rangle_{\rm t}$ (see Eq. 23) in the case of $B_{\rm 0y} = 30\;$G, $A=10\;\text{km}\cdot\text{s}^{-1}$, and $y_{\rm 0} = 0.3\;$Mm, and $w=0.2\;$Mm, with varying $B_{\rm 0z}$ in the range $0-10\;$G. Because of the much larger amplitude of the initial pulse, the obtained results are no longer insignificant. We note that the velocity values fall off with $B_{\rm 0z}$. For $B_{\rm 0z}=0\;$G, a down-flow occurs with its minimum velocity of about $-1.2\;$km$\cdot$s$^{-1}$ at $y\approx2\;$Mm, and at higher altitudes, for example\ at $y=20\;$Mm, an up-flow takes place with its maximum velocity of about $1.5\;$km$\cdot$s$^{-1}$. In the case of $B_{\rm 0z}=5\;$G, $\langle V_{\rm iy}\rangle_{\rm t}$ reveals a similar trend, as in the case of $B_{\rm 0z}=0\;$G, but it reaches smaller values: a down-flow minimum velocity is about $-1\;$km$\cdot$s$^{-1}$ at $y\approx2\;$Mm, and an up-flow maximum velocity at $y=20\;$Mm is $\approx1.1\;$km$\cdot$s$^{-1}$. For the maximum considered magnitude of the transverse magnetic field, $B_{\rm 0z}=10\;$G, down-flow occurs with its minimum velocity $\approx0.8\;$km$\cdot$s$^{-1}$, and an up-flow maximum velocity is $\approx0.7\;$km$\cdot$s$^{-1}$. This outcome can be compared to the results from \citet{2021A&A...652A.114P}, where an almost identical case to the one from the top panel is discussed; the only difference being that the current pulse width is larger. So, in that case (Fig.~5, bottom right panel), a down-flow takes place with minimum velocity $\approx-0.7\;$km$\cdot$s$^{-1}$ at height $y\approx2\;$Mm, and higher up, at about $y=5\;$Mm, an up-flow takes place with its maximum velocity of about $0.35\;$km$\cdot$s$^{-1}$. From this, it can be inferred that values that are almost twice larger the current values result from a pulse width that is twice larger. 
\begin{figure*}
    \begin{center}
        \hspace{-0.5cm}
        \includegraphics[width=0.38\textwidth]{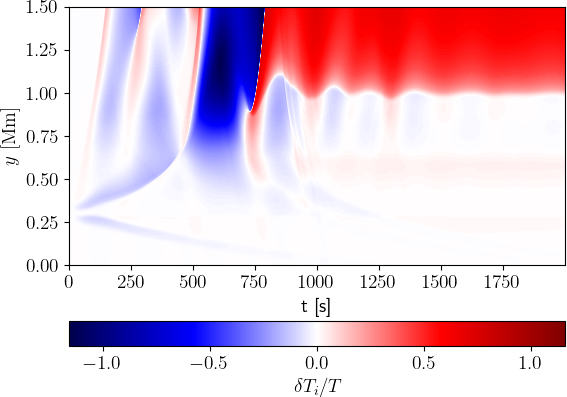}
        \hspace{1cm}
        \includegraphics[width=0.36\textwidth]{Viy_by30-bz0_y03_a10_w02.png}
        \hspace{1cm}
        \includegraphics[width=0.37\textwidth]{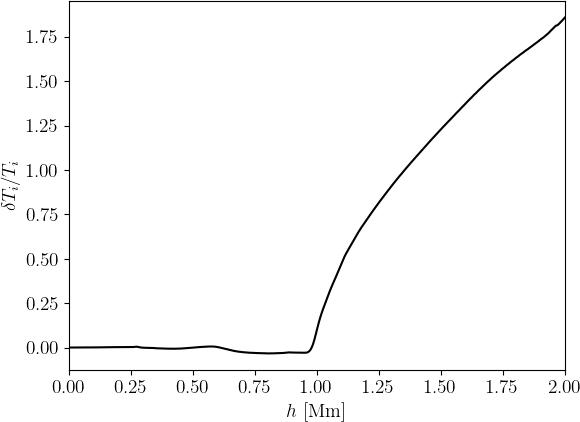}
        \hspace{0.6cm}
        \includegraphics[width=0.37\textwidth]{Ait_Viy_by30_Bz5_y03_A10.png}
        \vspace{-0.5cm}
    \end{center}
    \vspace{-0cm}
    \caption{Time-distance plots for $\delta T_{ie}/T$ (top left) and $V_{\rm iy}$ (top right), and $<\delta T_{ie}/T>_{\rm t}$ (bottom left) and $<V_{\rm iy}>_{\rm t}$ (bottom right) averaged over $x$ in the case of $B_{\rm 0y} = 30\;$G, $B_{\rm 0z} = 5\;$G, $y_{\rm 0} = 0.3\;$Mm,  $w=0.2\;$km, and $A=10\;\text{km}\cdot\text{s}^{-1}$.}
\end{figure*}

Figure~10 (top panels) presents time--distance plots for the perturbed ion temperature $\delta T_{\rm ie}/T$ (left panel) and for the vertical component of the ion velocity $V_{\rm iy}$ (right panel) in the case of $A=10\;\text{km}\cdot\text{s}^{-1}$ and $y_{\rm 0}=0.3\;$Mm, $B_{\rm 0y}=30\;$G and $B_{\rm 0z}=5\;$G. Comparing with the results from \citet{2021A&A...652A.114P}, it is discernible that the maximum value of the perturbed relative ion temperature is slightly higher here, with $\max(\delta T_{\rm ie}/T) \approx 1.3$, than in Paper~I, where $\max(\delta T_{\rm ie}/T) \approx 1$. The maximum value of the vertical component of ion velocity is also slightly higher; in this case max$(V_{\rm iy})\approx70\;\text{km}\cdot\text{s}^{-1}$ in the range of $0-20\;$Mm, while in the previous paper, max$(V_{\rm iy})\approx65\;\text{km}\cdot\text{s}^{-1}$.

The bottom panels of Figure~10 show the perturbed relative ion temperature $\delta T_{\rm ie}/T$ averaged over time, and the vertical component of the ion velocity $V_{\rm iy}$ averaged over time (see Eqs.~(22)-(23)). The trend in the plot from the left panel is similar to the same plot from Paper~I (the corresponding values are also equal), but the right plot differs. It is noticeable that slow down-flow takes place in the lower atmosphere, up to $y \approx 7\;$Mm. Higher up, an outflow takes place, with a magnitude that is growing with height $y$; it reaches about $1.1\;\text{km}\cdot\text{s}^{-1}$ at $y=20\;$Mm (this outflow velocity is three times smaller in Paper~I -- about $0.35\;\text{km}\cdot\text{s}^{-1}$). Below $y \approx 5\;$Mm, the down-flow in the present case reaches a minimum value of about $-1\;\text{km}\cdot\text{s}^{-1}$ (comparing to $-0.65\;\text{km}\cdot\text{s}^{-1}$). This means that the down-flow turns into an outflow when going from the lower atmosphere to the corona.
\begin{figure}
    \begin{center}
        \hspace{0cm}
        \includegraphics[width=0.45\textwidth]{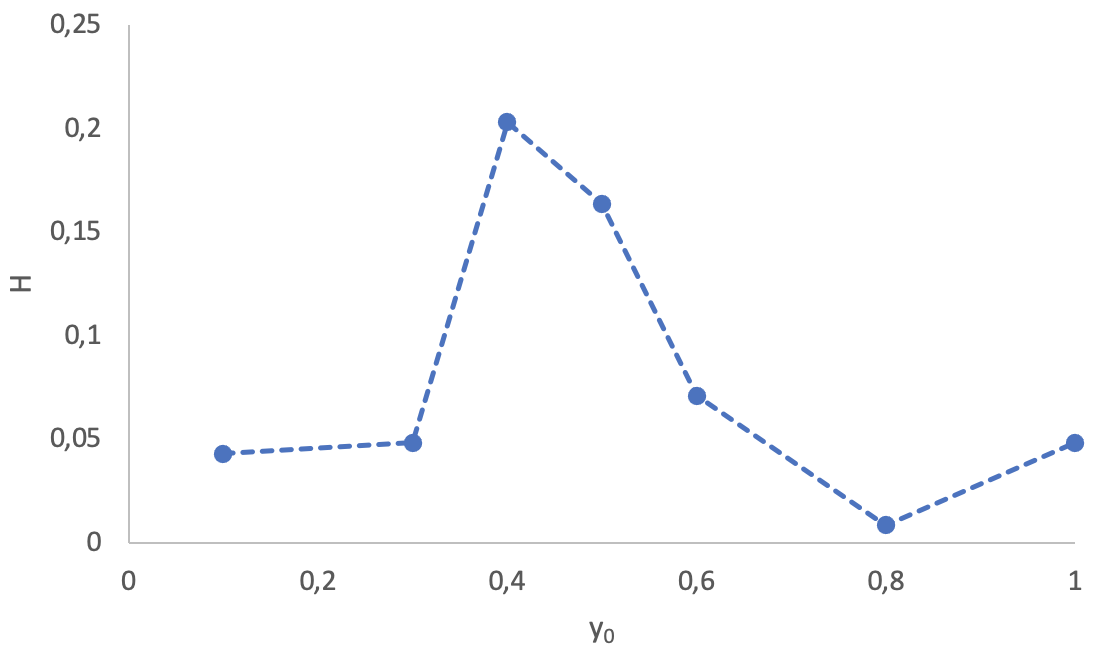}
        \vspace{0cm}
        \includegraphics[width=0.45\textwidth]{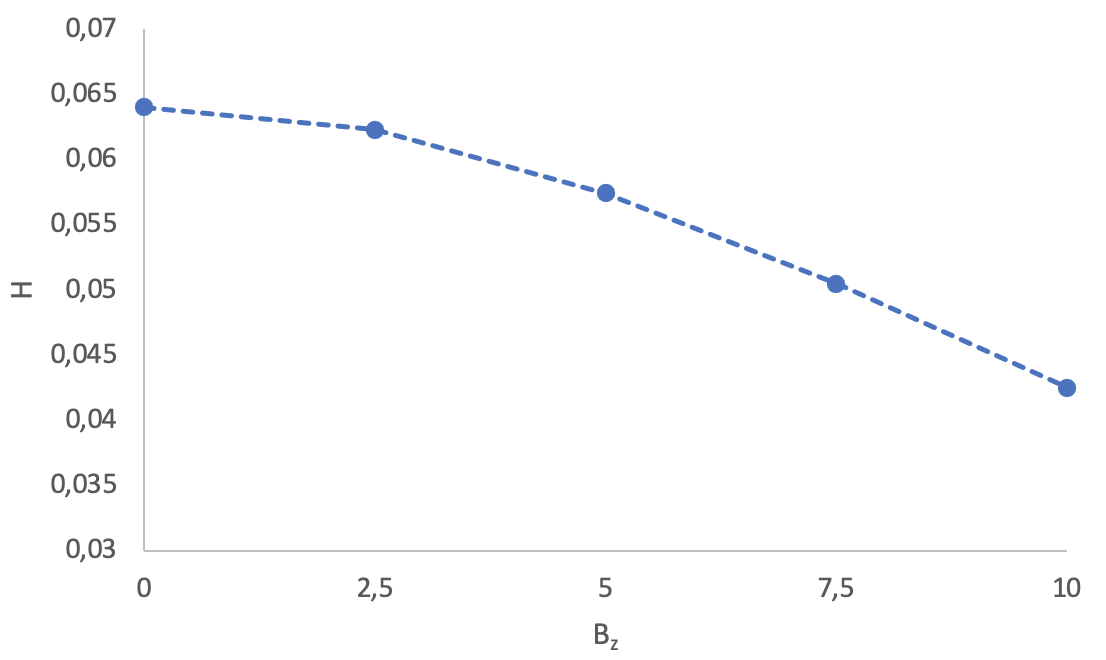}
        \vspace{0cm}
    \end{center}
    \vspace{-0.5cm}
    \caption{Relative perturbed temperature of ions averaged over time and height, $H$, vs. $y_{\rm 0}$ (top) and vs. $B_{\rm 0z}$ (bottom) for $B_{\rm 0y} = 30\;$G, $y_{\rm 0}=0.3\;$Mm, and $A = 10\;\text{km}\cdot\text{s}^{-1}$.}
\end{figure} 

Figure~11 displays the relative perturbed ion temperature averaged over height and time in the case of $B_{\rm 0y} = 30\;$G and $A = 10\;\text{km}\cdot\text{s}^{-1}$. This quantity is defined as
\begin{equation}
   H=\frac{1}{y_{\rm 1}-y_{\rm 0}}\int_{y_{\rm 0}}^{y_{\rm 1}}H(y)\, dy\, ,
\end{equation}
with
\begin{equation}
   H(y)=\frac{1}{t_{\rm 2}-t_{\rm 1}}\int_{t_{\rm 1}}^{t_{\rm 2}}\frac{\delta T_{\rm ie}}{T}\, dt\, ,
\end{equation}
where $t_{\rm 1}=0\;$s, $t_{\rm 2}=3000\;$s, $y_{\rm 0}$ is the initial pulse location, and $y_{\rm 1}=20\;$Mm. The top panel illustrates $H$ versus the launching height $y_{\rm 0}$. It is noticeable that the minimal heating occurs for $y_{\rm 0} = 0.8\;$Mm and the maximal heating $H$ is obtained for $y_{\rm 0} = 0.4\;$Mm, which is close to the top of the photosphere. It is noteworthy that $H$ drops with $y_{\rm 0}>0.4\;$Mm. This behavior can be explained by the fact that magnetoacoustic waves release thermal energy as a result of ion--neutral collisions. The magnitude of $H$ falls off with $y_{\rm 0}$ in the upper layers because the ionization degree grows with height.
The obtained values are almost twice smaller than the results from the previous paper (for $y_{\rm 0} = 0.4\;$Mm, $H$ is equal to 0.2, compared to $H=0.5$ for the same localization) because of the pulse width $w $ being twice smaller.

Figure~11 (bottom panel) illustrates the dependence of the average heating $H$ on the transversal magnetic field $B_{\rm z}$ for a pulse located at $y_{\rm 0} = 0.3\;$Mm. Here, it is clearly seen that the higher the value of a transversal magnetic field, the less heat is deposited in the atmosphere. This is understandable as for a higher value of $B_{\rm 0z}$, the coupling between Alfv\'en and magnetoacoustic waves becomes stronger. As a result, more energy is transferred to magnetoacoustic waves.
\section{Summary and conclusions}
In this paper, the results of numerical simulations of impulsively generated linearly coupled two-fluid Alfv\'en and magnetoacoustic waves are presented and discussed. Both waves are known to contribute to the heating of the solar  chromosphere and to the driving of plasma outflows \citep{2021A&A...652A.114P, 2021A&A...652A.124N}. Wave energy thermalization takes place in both cases as a result of ion-neutral collisions. We attempted to discover the effect of the wave dissipation on the ion temperature and on the generation of vertical plasma flows. In Paper~I, only Alfv\'en waves were examined. In the present paper, the focus was on the contribution of coupled Alfv\'en and magnetoacoustic waves. All simulations were performed using the JOANNA code \citep{2020A&A...635A..28W} on the basis of the two-fluid model. 

Two values of initial pulse amplitude $A$, mainly  $A=1\;\text{km}\cdot\text{s}^{-1}$ and $A=10\;\text{km}\cdot\text{s}^{-1}$, were used to investigate the problem. The obtained results indicate that the Alfv\'en and magnetoacoustic waves (both alone and coupled) that are generated in the middle of the photosphere at $y_{\rm 0}=0.3\;$Mm with a small initial amplitude, negligibly contribute to the thermal energy of the system, and this can only slightly accelerate the plasma outflows. On the other hand, initial pulses with a much larger initial amplitude, which are still physically feasible and realistic, can contribute more substantially to chromosphere heating. The obtained results can be compared to the work by \citet{2018NatPh..14..480G}, in which the first observational evidence of Alfv\'en wave dissipation in the chromosphere was found. The most probable cause of wave dissipation are ion-neutral collisions. \citet{2018NatPh..14..480G} revealed that the large amplitude Alfv\'en waves cause an increase in temperature up to 5\%. Moreover, these higher amplitude waves can also result in plasma outflows, which can become more considerable as they rise in altitude and eventually become the potential source of the solar wind. These results are in a agreement with the findings of Paper~I and of \citet{2021A&A...652A.124N}. 

In addition, it was found that for a vertical magnetic field of $B_{\rm 0y} = 30\;$G, 
the value of the transverse magnetic field component, $B_{\rm 0z}$, plays a significant role in the system evolution. 
A nonzero $B_{\rm 0z}$ component indicates the existence of magnetoacoustic waves. A higher value of the $B_{\rm 0z}$ component results in a higher plasma outflow velocity, but in a slight decrease in temperature.
The cases with the higher amplitude, mainly the case of $A=10\;\text{km}\cdot\text{s}^{-1}$ and with the transverse magnetic field component of $B_{\rm 0z} = 10\;$G, reveal that the coupled waves heat the chromosphere  more significantly  and also accelerate the plasma more. 
It was found that the maximum heating corresponds to the pulse that was initially launched from the middle of the photosphere ($y_{\rm 0}=0.3\;$Mm), which is about $200-300\;$km below the temperature minimum height. The magnitude of the flows, however, was found to be small and substantially lower than the observed inflows 
and outflows.

In summary, from the obtained results in the present paper, it can be concluded that the initial pulse amplitude plays a significant role in the variation of the heating degree and the magnitude of the generated plasma outflows. When changing the pulse width from $0.05\;$Mm to $0.2\;$Mm, the relative temperature increases approximately seven times. However, the maximum velocity in both the vertical and transversal components decreases by about three to four times. Unfortunately, the numerical results do not fully fit to the observational data, even though the obtained flow amplitudes are in the observed ranges, and therefore further investigations are required.

\section*{Acknowledgements}
The JOANNA code has been developed by Darek Wójcik with 
some contribution from 
Luis Kadowaki 
and 
Piotr Wołoszkiewicz. 
This work was done within the framework of the projects from the Polish National Foundation (NCN) grant No. 2020/37/B/ST9/00184. Numerical simulations were performed on the MIRANDA cluster at Institute of Mathematics of University of M.\ Curie-Skłodowska, Lublin, Poland. 

\bibliographystyle{aa}
\bibliography{bibliography.bib}
\end{document}